\documentclass[10pt]{article}

\usepackage{amsfonts}
\usepackage{amssymb}
\usepackage{natbib}
\newcommand{\tr}{\, \mbox{Tr}\, }

\newcounter{thaler}
\newenvironment{mlist}{\begin{list}{\arabic{thaler}}%
{\usecounter{thaler}
\setlength{\rightmargin}{\leftmargin}
\topsep=0pt
\itemsep=0pt
\parskip=0pt
\parsep=0pt
}}{\end{list}}

{\begin{center}%
\begin{tabular}{|c|} \hline \\ $\displaystyle \ \ \ }%
{\ \ \ $\\ \\ \hline \end{tabular} \end{center}}

\newcounter{thingy}
\renewcommand{\thethingy}{\arabic{section}.\arabic{thingy} }

\renewcommand{\section}[1]{ \refstepcounter{section}
\setcounter{thingy}{0}
{\bf \arabic{section}.~#1 }}{}

{}

\newenvironment{definition}
{ \refstepcounter{thingy}{\bf \thethingy Definition} }{}

\newenvironment{theorem}
{ \refstepcounter{thingy}{\bf \thethingy Theorem} \em }{}

\newenvironment{proposition}
{ \refstepcounter{thingy}{\bf \thethingy Proposition} \em }{}

\newenvironment{problem}
{ \refstepcounter{thingy}{\bf \thethingy Problem} \em }{}

\newcommand{\ket}[1]{| #1 \rangle}
\newcommand{\bra}[1]{\langle #1 |}
\newcommand{\proj}[1]{\ket{#1}\! \bra{#1}}

\newcommand{\text}{\mbox}

\newcommand{\trace}{\tr}

\newcommand{\real}{\mathrm{Re}}

\newcommand{\ketbra}[2]{\ket{#1}\bra{#2}}
\newcommand{\lie}[1]{\mathfrak{#1}}

\newcommand{\trless}[1]{{#1}_{\!\circ}}

\def\openone{\leavevmode\hbox{\small1\kern-3.8pt\normalsize1}}
\def\RR{{\rm I\kern-.2emR}}
\def\tr{{\rm Tr}\; }
\def\ce{{\cal E}}
\def\cl{{\cal L}}

\def\cb{{\cal B}}

\def\ch{{\cal H}}

\def\ca{{\cal A}}
\def\cR{{\cal R}}

\def\cz{{\cal Z}}

\renewcommand{\trace}{\mbox{tr}}

\def\intersect{\cap}

\def\extr {\rm Extr~}

\def\dr{\smash{\downharpoonright}}

\def\R{{\mathbb{R}}}

\def\openone{\leavevmode\hbox{\small1\kern-3.8pt\normalsize1}}
\def\RR{{\rm I\kern-.2emR}}
\def\tr{{\rm tr}\; }

\def\fg{\mathfrak{g}}
\def\fh{\mathfrak{h}}

\def\fsu{\mathfrak{su}}

\def\one{{\mathchoice {\rm 1\mskip-4mu l} {\rm 1\mskip-4mu l} {\rm
1\mskip-4.5mu l} {\rm 1\mskip-5mu l}}}

\newcommand{\Id}{\text{id}}
\newcommand{\dmelement}[2]{ \langle #1 | #2 | #1 \rangle}
\newcommand{\beq}{\begin{equation}}
\newcommand{\eeq}{\end{equation}}
\newcommand{\beqa}{\begin{eqnarray}}
\newcommand{\eeqa}{\end{eqnarray}}
\begin{document}

\bibliographystyle{apsrmp}

\begin{center}
{\large \bf Generalization of entanglement to convex operational theories:
Entanglement relative to a subspace of observables} 

Howard Barnum\footnote{\ Los Alamos National Laboratory, 
Los Alamos, NM 87545\\
%\hspace*{3.8mm}$\mbox{}^2$ National Institute of Standards and Technology,
%Boulder, CO 80305. \\ 
\hspace*{3.8mm}$\mbox{}^2$ Department of Physics and Astronomy, 
Dartmouth College, 6127 Wilder Laboratory, \hspace*{6.7mm}Hanover, NH 03755}, 
%Emanuel Knill$\mbox{}^2$, 
Gerardo Ortiz$\mbox{}^1$, Rolando Somma$\mbox{}^1$, 
Lorenza Viola$\mbox{}^2$ 
%\vspace*{1mm}

%\sl Submitted to Proceedings of Quantum Structures VII, \\
%pecial Issue of International Journal of Theoretical Physics}
\end{center} 

\begin{abstract}
We define what it means for a state in a convex cone of states on a
space of observables to be {\em generalized-entangled} relative to a
subspace of the observables, in a general ordered linear spaces
framework for operational theories.  This extends the notion of
ordinary entanglement in quantum information theory to a much more
general framework.  Some important special cases are described, in
which the distinguished observables are subspaces of the observables
of a quantum system, leading to results like the identification of
generalized unentangled states with Lie-group-theoretic coherent
states when the special observables form an irreducibly represented
Lie algebra.  Some open problems, including that of generalizing the
semigroup of local operations with classical communication to the
convex cones setting, are discussed\vspace*{2mm}.  \\ KEY WORDS:
Entanglement, convexity, ordered linear spaces, operational theories, 
observables.  
\\ PACS NUMBERS: 03.65.Ud
\end{abstract} 

\iffalse
\underline{\bf Outline}
\begin{mlist}
\item[(1)] Introduction
\item[(2)] States and weights on test spaces
\item[(3)] States on coupled systems
\item[(4)] Operator representations of local states
\item[(5)] Decomposable states 
\item[(6)] Bell inequalities 
\item[(7)] Teleportation
\item[(A)] Appendix: Background on tensors and operators.
\end{mlist}
\fi
   
\large    
\section{INTRODUCTION}
\normalsize

Entanglement is a distinctively quantum phenomenon whereby a pure
state of a composite quantum system may no longer be determined by the
states of its constituent subsystems~\citep{Schroedinger35a}. Entangled
pure states are those that have {\it mixed} subsystem states. To
determine an entangled state requires knowledge of the correlations
between the subsystems. As no pure state of a classical system can be
correlated, such correlations are intrinsically non-classical, as
strikingly manifested by the possibility of violating local realism
and Bell's inequalities~\cite{Bell93a}. In the science of quantum
information processing (QIP), entanglement is regarded as the defining
resource for quantum communication, as well as an essential feature
needed for unlocking the power of quantum computation.

The standard definition of quantum entanglement requires a preferred
partition of the overall system into subsystems--- that is,
mathematically, a factorization of the Hilbert space as a tensor
product.  Even within quantum mechanics, there are motivations for
going beyond such subsystem-based notions of entanglement.  Whenever
indistinguishable particles are sufficiently close to each other,
quantum statistics forces the accessible state space to be a proper
subspace of the full tensor product space, and exchange correlations
arise that are not a usable resource in the conventional QIP
sense. Thus, the natural identification of particles with preferred
subsystems becomes problematic.  Even if a distinguishable-subsystem
structure may be associated to degrees of freedom different from the
original particles (such as a set of position or momentum modes, as in
~\cite{Zanardi2001a}), inequivalent factorizations may occur on the
same footing.  Entanglement-like notions not tied to modes have been
proposed for bosons and fermions~\citep{Eckert2002a}. 
However, the introduction of quasiparticles, or the purposeful
transformation of the algebraic language used to analyze the system
~\citep{Batista2001a, Batista2002a}, may further complicate the
choice of preferred subsystems.

In this paper, we review and further develop \emph{generalized
entanglement} (GE) introduced in~\cite{BKOV2002a}, which incorporates
the entanglement settings introduced to date in a unifying framework.
In quantum-mechanical settings, the key idea behind GE is that
entanglement is an \emph{observer-dependent concept}, whose properties
are determined by the expectations of a \emph{distinguished subspace
of observables} of the system of interest, without reference to a
preferred subsystem decomposition.  Distinguished observables may
represent, for instance, a limited means of manipulating and measuring
the system.  Standard entanglement is recovered when these means are
restricted to arbitrary \emph{local} observables acting on subsystems.
The central idea is to generalize the observation that standard
entangled pure states are precisely those that look mixed to local
observers.

The most fundamental aspects of this notion of GE make use only of the
convex structures of the spaces of quantum states and observables.
Therfore it is also applicable in contexts much broader than that of
quantum systems with distinguished subspaces of observables.  It may
be formulated within general convex frameworks, based on ordered
linear spaces or the closely related notion of convex effect algebras,
suitable for investigating the foundations of quantum mechanics and
related physical theories (cfr.~\cite{Beltrametti97a} and references
therein).  While commenting on physically motivated special cases, we
will concentrate on this general setting in the present paper.  Though
we make no major advances over \cite{BKOV2002a} and \cite{BKOSV2004a},
new material here  includes
Theorem 3.4 which gives another characterization of the convex cones
framework, in terms of restriction to a subspace of observables, and 
more detailed investigation of the distinguished quantum observables
subspace.  This includes the introduction of the {\em unique preimage property}
(Def. \ref{def: unique preimage property}) and the relationship between
the quadratic purity measure, generalized entanglement, and the UPIP in
this context, notably Problems \hbox{\ref{prob: when UPIP},} 
\ref{prob: when GE implies mrp}, and \ref{prob: irreducible implies UPIP},
and Propositions \ref{prop: when else mrp implies ge} and 
\ref{prop: max purity implies GE}.

Two sets of articles contain related
ideas.  The first originated in the context of $C^{*}$ and von Neumann
algebras, for example in~\cite{CNT87a}, where the dynamical entropy of
automorphisms of algebras, intended to generalize the Kolmogorov-Sinai
dynamical entropy, is defined --- using a notion of entropy of a
state's restriction to a subalgebra introduced in~\cite{NT85a}.  These
ideas were further developed with special attention to finding optimal
decompositions for the convex roof construction of entropy relative to
a subalgebra, and applied to quantum information concepts such as
quantum parameter estimation and the entanglement of formation.  See
e.g.~\cite{Benatti96a, Uhlmann98a,BNU96a,BN98a,BNU2003a}.  The
association of subsystems, whether physical or
``virtual''/''encoded,'' of a quantum system with associative
subalgebras appeared in in a second set of
articles~\cite{KLV2000a,DeFilippo2000a,VKL2001a,Zanardi2001b}; this
association was recently revisited, and examples collected,
%% L: Perhaps 'reviewed' might be a bit too strong?... 
in~\cite{ZLL2004a}.  Note, however, that these latter articles were
not directly concerned with the {\em extremality properties of reduced
states} which form the basis of our GE notion.  Also, in both sets of
articles, the context of subalgebras, whether $C^*$, von Neumann, or
associative, is considerably more restrictive than the general context
we work in here, except for the fact that Benatti, Connes, Narnhofer,
Thirring, and Uhlmann often include and are sometimes primarily
interested in infinite-dimensional algebras, whereas we focus here
exclusively on the finite-dimensional setting.
%% to avoid technical issues.  L: Do we really have to say that?

\iffalse 
Here we highlight the significance of GE from a physics and
information-physics perspective. For this purpose, we focus on the
case where the observable subspace is a Lie algebra. A key result is
then the identification of pure generalized unentangled states with
the \emph{generalized coherent states} (GCSs, a connection
independently noted by Klyachko~\cite{klyachko2002a}), which are well
known for their applications in physics~\cite{Zhang90a}.    
We demonstrate that many information-theoretic notions previously thought
specific to partitioning into subsystems extend to coherent state 
theory and beyond, define new measures of entanglement based on the 
general theory, and apply quantum information to condensed-matter problems. 
In particular, we introduce notions of \emph{Generalized Local Operations 
assisted by Classical Communication} (GLOCC) under which the ordinary 
measures of standard entanglement do not increase, as well as measures 
of GE with the desired behavior under classes of GLOCC maps. New measures 
of standard entanglement can be constructed for the multipartite case. 
In the Lie-algebraic setting, a simple GE measure obtained from the
\emph{purity relative to a Lie algebra} is a useful diagnostic tool
for quantum many-body systems, playing the role of a \emph{disorder
parameter} for broken-symmetry quantum phase transitions.
\fi

\large 
\section{MATHEMATICAL BACKGROUND}
\normalsize

For background on cones and convexity, we highly recommend the
text by~\cite{Barvinok2002a}, or the short introductory portion
of~\cite{HHL89a}; however, the summary we give here should suffice for
what follows.

\begin{definition}
A {\em positive cone} is a proper subset $K$ of a real vector space
$V$ closed under multiplication by nonnegative scalars.  It is called
{\em regular} if it is (a) convex (equivalently, closed under
addition: $K + K = K$), (b) generating ($K-K=V$, equivalently $K$
linearly generates $V$,) (c) pointed ($K \cap -K = \{0\}$, so that it
contains no non-null subspace of $V$), and (d) topologically closed
(in the Euclidean metric topology, for finite dimension).
\end{definition}
In the remainder of this paper, ``vector space'' and ``linear space''
will mean {\em finite-dimensional} vector space, ``cone'' will mean a
regular cone in a finite-dimensional vector space, unless otherwise
stated.  

A cone $K$ induces a partial order $\ge_K$ on $V$, defined by
$x \ge_K y := x - y \in K$.  \iffalse $(V,\ge_K)$, or sometimes
$(V,K)$, is called an {\em ordered linear space}.  \fi \iffalse The
Hermitian operators on a finite-dimensional complex vector space, with
the ordering induced by the cone of positive semidefinite operators, 
are an example.  \fi \iffalse (A relation $R$ is defined to be
a partial order if it is reflexive ($x R x$), transitive ($x R y ~\&~
y R z \Rightarrow x R z$) and antisymmetric ($(x R y ~\&~ y R x)
\Rightarrow x=y$).)  \fi It is ``linear-compatible'': inequalities can
be added, and multiplied by positive scalars.  If one removes the
requirement that the cones be generating, cones are in one-to-one
correspondence with linear-compatible partial orderings.  A pair
$\langle V, \succeq \rangle$ of a linear space and a distinguished
such ordering is called an {\em ordered linear space}.  The categories
of real linear spaces with distinguished cones and partially ordered real
linear spaces are equivalent.

Note that the intersection of the interior of a generating cone with a
subspace is (if not equal to $\{0\}$) a ({\em non-closed} but otherwise 
regular) cone that
generates the subspace.  When a cone or other set is said to generate
a linear space, it does so via linear combination.  When a set is said
to generate a cone, it does so via positive linear combination.  We
will use the notation $\dot{C}$, for the set $C - \{0\}$.

By an {\em extremal} state in a convex set of states, we mean the
usual convex-set notion that a point $x$ is extremal in a convex set
$S$ if (and only if) it cannot be written as a nontrivial convex
combination $x = \lambda_1 x_1 + \lambda_2 x_2$ of points $x_1, x_2$
in $S$.  (Convex combination means $\lambda_i
\ge 0, \lambda_1 + \lambda_2 = 1$, and nontrivial means $\lambda_i \ne
0, x_1 \ne x_2$).  We sometimes use the physics term {\em pure state}
for an extremal point in a convex set of states, but for clarification
we emphasize that when this convex set is the set of all quantum
states on some Hilbert space, the term ``pure state'' in the present
paper refers to a projector $\pi := \proj{\psi}$, and not to a vector
$\ket{\psi}$ in the underlying Hilbert space.  We write $\extr{S}$ for
the set of extremal points of a convex set $S$.

A {\em ray} belonging to a cone $K$ is a set $R$ such that there
exists an $x \in K$ for which $R = \{\lambda x: \lambda \ge 0 \}$,
i.e. it is the set of all nonnegative scalar multiples of some element
of the cone.  An {\em extreme ray} in $K$ is a ray $R$ such that no $y
\in R$ can be written as a convex (or equivalently, positive)
combination of elements of $K$ that are not in $R$.  The topological
closure condition guarantees, through an easy but not trivial argument
using the Krein-Milman theorem, that a (regular) cone is convexly 
(equivalently, positively)
generated by its extreme rays.    We'll say a point is
extremal in a cone if it belongs to an extreme ray of the cone;  note
that such points are not usually extremal in the convex set sense,
although the cone is a convex set;  the only point in a cone extremal
in the convex set sense is zero.

%Duality is often a useful tool when dealing with cones.  
The dual vector space $V^*$ for real $V$ is the space of all linear
functionals from $V$ to $\R$; the dual cone $C^* \subset V^*$ of the
cone $C \subset V$ is the set of such linear functionals which are
nonnegative on $C$.
% For finite dimensional vector spaces, 
% $V^*$ is isomorphic to $V$ as a vector space (they
% have the same dimension).  However, this isomorphism is not 
% canonical; any 
% nonsingular linear map from $V$ onto $V^*$ does the job.  
$\lambda \in  V^*$ is said to
separate $C$ from $-C$ if $\lambda(x) \ge 0$ for all nonzero $x \in
C$.  For $\alpha \in V^*$, $x
\in V$, we write the value of $\alpha$ on $x$ as $\alpha[x]$, rather
than $\alpha(x)$.
% The action of
% a map $\phi$ on a linear space may be written either in standard
% functional notation ($x \mapsto \phi(x)$, or, as in the next
% definition, via superscripting: $x \mapsto x^\phi$.  
The adjoint $\phi^*: V_2^* \rightarrow V_1^*$ of a linear map $\phi:
V_1 \rightarrow V_2$ is defined by \beqa \phi^*(\alpha)[x] =
\alpha[\phi(x)]\;, \eeqa for all $\alpha \in V_2^*, x \in V_1$.  The
following proposition is easily (but instructively) verified.

\begin{proposition} \label{dual maps}
Let $C_i$ be a cone in $V_i$ for $i = 1,2,$ and let $\phi(C_1)
\subseteq C_2$.  Then $\phi^*(C_2^*) \subseteq C_1^*$.
\end{proposition}

We will also use the following:

\begin{proposition} \label{epimono}
Let $C_i$ be a cone in $V_i$ for $i = 1,2,$ and let $\phi(C_1) = C_2$.
Then $\phi^*(C_2^*) \subseteq C_1^*$ and $\phi^*$ is one-to-one.
\end{proposition}

{\bf Proof:} 
Let $\eta_1, \eta_2 \in C_2^*$, and  $\eta_1 \ne eta_2$.  $\eta_1 \ne \eta_2$ is
equivalent to the existence of $y$ in $C_2$ such that $\eta_1[y] \ne
\eta_2[y]$.  By the assumption that $\phi$ maps $C_1$ onto $C_2$,
there is an $x \in C_1$ such that $\phi(x)=y$; thus $\eta_1[\phi(x)]
\ne \eta_2[\phi(x)]$.  By the definition of $\phi^*$, this implies
that $\phi^*(\eta_1)[x] \ne \phi^*(\eta_2)[x]$, which implies that
$\phi^*(\eta_1) \ne \phi^*(\eta_2)$.  $\Box$

\large
\section{GENERALIZED ENTANGLEMENT}
\normalsize

We now introduce GE of states in a convex set of states given by the
intersection $\hat{C}$ of an affine ``normalization'' plane $\{x :
\lambda(x) = \alpha \}$ (for a fixed $\alpha$, which we'll take to be
one) with a cone $C$ of ``unnormalized states.''  This GE is a
relative notion: states are entangled or unentangled relative to
another such state-set $\hat{D}$, and a choice of
normalization-preserving map of the first state-set onto the second,
which generalizes the notion of computing the reduced density matrices
of a bipartite system.  To fix ideas, note that in the case where $C$
is supposed to represent states on a finite dimensional quantum system
whose Hilbert space has dimension $d$, $C$ is isomorphic to the set of
$d \times d$ positive semidefinite matrices, whose normalized (i.e.
unit-trace) members form the convex set of density matrices for the
system, while the ambient linear space $V$ is the space of $d \times
d$ Hermitian matrices.  We shall often use the abbreviation ``PSD''
for ``positive semidefinite.''

\begin{definition} \label{def: normalized unentangled} 
Let $V,W$ be finite-dimensional real linear spaces equipped with cones
$C \subset V$, $D \subset W$, and distinguished linear functionals
$\lambda \in C^*$, $\tilde{\lambda} \in D^*$ that separate $C,D$ from
$-C, -D$ respectively.  Let $\pi: V \rightarrow W$ be a linear map that
takes $C$ onto $D$ (that is, $\pi(C) = D$), and maps the affine plane
$L_\lambda := \{x \in V : \lambda(x)=1\}$ onto the plane
$M_{\tilde{\lambda}} := \{y \in W : \tilde{\lambda}(y) = 1 \}$.  
An element (``state'') in $\hat{C} := L_\lambda \intersect C$ is called
{\em generalized unentangled} (GUE) relative to $D$ if it is in the
closure of the convex hull of the set of extreme points $x$ of $\hat{C}$
whose images $\pi(x)$ are extreme in 
$\hat{D} := D \intersect M_{\tilde{\lambda}}$.
\end{definition}

\begin{definition}\label{def: cone-pair}
We will call a pair of linear spaces $V,W$ equipped
with distinguished cones $C,D$, functionals $\lambda,
\tilde{\lambda}$, and a map $\pi$, satisfying the conditions in the
above definition, a {\em cone-pair}.  As noted above, we write $\hat{C}, \hat{D}$ for
the normalized subsets of $C, D$, i.e. for $\{x \in C: \lambda(x) =
1\}$ and $\{x \in D: \tilde{\lambda}(x) = 1\}$.  We will also
sometimes call $\lambda, \tilde{\lambda}$ the {\em traces} on their
respective cones, so that the condition on $\pi$ above may be called
{\em trace-preservation}.
\end{definition}
That is, with the usual physics terminology that extremal states are
``pure'' and nonextremal ones ``mixed,'' unentangled pure states of
$\hat{C}$ are those whose ``reduced'' states (images under $\pi$) are pure,
and the notion extends to mixed states as in standard entanglement
theory: unentangled mixed states in $\hat{C}$ are those expressible as
convex combinations of unentangled pure states (or limits of such
combinations, though the latter is unnecessary in finite dimension).

It is easy to see that the motivating example of ordinary bipartite
entanglement is a special case of this definition.  Here, $C$ is the
cone of PSD operators on some tensor product $A \otimes B$ of
finite-dimensional Hilbert spaces, while $D$ is the direct product of
the cones of PSD operators on $A$ and on $B$ (intuitively, it is just
the cone of all ordered pairs whose first member is a positive
operator on $A$ and whose second is one on $B$).  $\lambda$ is 
the trace.  $\pi$ is the map
that takes an operator on $A \otimes B$ to the ordered pair of its
``marginal'' or ``reduced'' operators (``partial traces'') on $A$ and
$B$.  Similarly, standard multipartite entanglement is a special case of GE.  
So we may view the GUE definition (in particular condition (a) of
Definition 3.3 below) as based on extending the long-standing
observation that for ordinary multipartite finite-dimensional quantum
systems, a pure state is entangled if and only if at least one of its
reduced density matrices is mixed.

It is perhaps mathematically more natural to define the {\em
unnormalized} unentangled states of $C$ relative to $D$, omitting all
mention of $\lambda, \tilde{\lambda}$, and the
normalization-preservation requirement on $\pi$.  That is:

\begin{definition}
Let $C, D$ be cones in finite-dimensional real linear spaces 
$V,W$ respectively, and let $\pi: V \rightarrow W$ map $C$
onto $D$.  $x \in C$ is {\em generalized unentangled} (relative to
$D, \pi$) if either (a) $x$ belongs to an extreme ray of $C$, and
$\pi(x)$ belongs to an extreme ray of $D$, or (b) $x$ is a positive
linear combination of states satisfying (a), or a limit of such
combinations.
\end{definition}

It is easy to verify that the unnormalized GUE states are a (possibly
non-generating, but otherwise regular) cone in $V$.  If one introduces
the notion of normalization in $C$ via a functional $\lambda$, it is
also easily verified that the normalized GUE states of Definition
\ref{def: normalized unentangled} are precisely the intersection of
this cone with the normalization plane.  (It is straightforward
to introduce a normalization plane, and associated functional
$\tilde{\lambda}$ on $W$, if desired, as the image of $L_\lambda$
under $\pi$.)

\cite{BKOV2002a}, and especially~\cite{BKOSV2004a},
stressed applications in which the reduced state-set is obtained by
selecting a distinguished subspace of the observables (Hermitian
operators) on some quantum system.  The reduced state-set is then the
set of linear functionals (equivalently, consistent lists of
expectation values for the distinguished observables) on this subspace
of the space of all observables, that are induced by normalized quantum
states\footnote{It is worth noting that beyond the setting of standard
quantum entanglement this is not in general a vacuous requirement:
there can be normalized linear functionals on the reduced state set
that are {\em not} obtainable by restriction from a quantum state on
the set of all observables. Although all normalized functionals on the
distinguished observables can be extended in many ways to normalized
functionals on the full set, in some cases not all can be extended to
{\em positive} functionals.}.  We dub this class of cone-pairs the {\em
distinguished quantum observables} setting.  Even in
the more general cones setting, there is a natural notion of
observables, and 
Definition~\ref{def: normalized unentangled} can  be interpreted
as restriction of the states to a subspace of the observables.  To show
this we employ a formalism of states, measurements, and observables
that, in many variants, is frequently used as a touchstone of
``operational'' approaches to theories in the abstract~\footnote{By an
``operational theory,'' we mean one in which a theory describes
various measurements or operations one can perform on systems of the
type described by the theory, and specifies a set of possible
``states,'' each of which determines the probabilities for the
outcomes of all possible measurements, when the system is in that
state.}.

We view $V^*$ as a space of real-valued observables.  For $x \in V^*$
and $\eta \in \hat{C}$, we interpret $x[\eta]$ as the expectation
value of observable $x$ in state $ \eta$.  We view $V$ as the dual of
$V^*$ in such a way that $x[\eta] = \eta[x]$ for all $x \in V^*, \eta
\in V$.  But what guarantee do we have that these expectation values
behave in a reasonable way, as observables in an operational theory
should?  That is, can we view the expectation value $\eta(x)$ of an
observable $x$ in a state $\eta$ as the expected value of some
quantity being measured?  By this we mean that $x$ is associated with
a quantity that takes different values depending on the outcome of the
measurement, and the state determines the expectation value by
determining probabilities for the different outcomes of the
measurement, such that the value $\eta(x)$ is indeed the expectation
value of the outcome-dependent quantity, calculated according to the
probabilities assigned to the outcomes by the state.

We will only sketch the answer to this question; more details may be
found in many places (though accompanied by additional concepts and
formalism), notably \cite{Beltrametti97a}.  In the structure we have
described, of state-space and dual observable space, we are able to
find a special class of observables, the ``decision effects,'' whose
expectation value may be viewed as the probability of a measurement
outcome.  These ``effects'' are the elements of the initial interval
$\ce := [0, \lambda] \subset C^*$, i.e.  the set of $x \in C^*$
satisfying \mbox{$\lambda \ge_{C^*} x$}.  A (finite) {\em resolution
of $\lambda$} is a set of effects $x_i \in \ce$ such that $\sum_i x_i
= \lambda$.  For normalized states $\omega$, it follows that
$\omega(x_i) \ge 0$ and $\sum_i \omega(x_i)=1$, so the values
$\omega(x_i)$ may be viewed as probabilities of measurement outcomes,
with a resolution of $\lambda$ representing the mutually exclusive and
exhaustive outcomes of some measurement.  Then it can be shown that
for {\em any} observable $A \in V^*$, a resolution $\cR$ of $\lambda$
and an assignment of real values $v(x_i)$ to the outcomes $x_i$ in $\cR$ can
be found, such that for all normalized states $\omega$, $\omega(A) =
\sum_i \omega(x_i) v(x_i)$.  For example, this is a consequence of (i)
of Theorem 1 in \cite{Beltrametti97a}.  In general the converse does
not hold, giving rise to a generalization of observables sometimes
known as {\em stochastic observables} for which not only does the
analogous statement (which is (i) of Theorem 1 of
\cite{Beltrametti97a} where stochastic observables are just called observables) hold,
but so does the converse of this analogue.  The relation between the
convex and the effect-algebras approach has been treated in various
places (and aspects of it appear in some contexts, e.g. \cite{Ludwig83a},
even earlier than the
formal notion of effect algebra).  Some references are
\cite{Gudder98a}, \cite{Gudder99a}, \cite{Gudder99b}, and the book
\cite{DallaChiara2004a} (especially Ch. 6).  \cite{Barnum2003a}
explores the relation between probabilistic operational theories and
``weak effect algebras,'' as well as related more dynamical objects termed {\em
operation algebras}, but without explicit consideration of observables.  
\cite{Bennett97a}, \cite{Foulis98a},  and \cite{Foulis2000a} address 
very closely related representational issues 
but without the constraint of convexity.  The relations 
between convex and general effect-algebras and their representations
are discussed in \cite{Foulis2005a}.

We now show that our formalism of maps $\pi$ onto cones $D$ is
equivalent to restriction to a subspace of observables.

\begin{theorem}\label{observable restriction equivalent to conepair}

I) (``Observable restriction implies cone-pair''). 
Let $C$ be a cone in $V$, and let $\lambda \in V^*$ separate $V$ from 
$-V$ (as in Def. \ref{def: normalized unentangled}), 
and let $W^*$ be a subspace of $V^*$, containing $\lambda$.  
For $\eta \in V$, define $\eta\dr: W^* \rightarrow
\R$ as the restriction of $\eta$ to the subspace $W^*$,
i.e. $\eta\dr(x) = \eta(x)$ for $x \in W^*$ and otherwise $\eta\dr(x)$
is undefined.  Thus $\eta\dr \in (W^*)^* =: W$.  
Define $D = \{\eta\dr : \eta \in C \}$, $M_\lambda = \{y
\in W : \lambda(y)=1\}$.
Define $\pi$ as the 
restriction map $\pi := \dr: V \rightarrow W, \eta \mapsto \eta\dr$.
Then $V, W, C,D, \lambda, \tilde{\lambda} (:= \lambda), \pi$
form a cone-pair in the sense of Definition \ref{def: cone-pair}.  
That is, $D$ is a cone in $W$, $\pi(C)=D$, and 
the image under $\dr$ of the plane $L_\lambda \equiv \{\eta \in V:
\lambda(\eta) = 1\}$ is a translation of a plane separating $D$ from
$-D$.

II) (``Cone-pair implies observable restriction'').
Let $V,W,C,D,\lambda, \tilde{\lambda},\pi$ be a cone-pair.  Then there
exists an injection (one-to-one map) $\tau: W^* \rightarrow V^*$, taking
$\tilde{\lambda}$ to $\lambda$, such that $\pi$ is the map from $V$ to $W$
that takes $x$ to the function $x\dr_{W^*}$.  Here $x \dr_{W^*}$ defined
as the 
linear functional on $W^*$ whose value on $a \in W^*$ is the value of 
$x$'s restriction to $\tau(W^*)$ on $\tau(a)$.  

\end{theorem}

{\em Remark concerning I:} The restriction that the subspace $W^*$ contain
$\lambda$ is hardly objectionable from an operational point of view.
$\lambda$'s expectation value is just the normalization constant, and
is independent of which normalized state has been prepared.  Therefore
it can be measured without any resources, and there is no point in
claiming that omitting it could represent a physically significant
restriction on the means available to observe or manipulate a system.

{\em Remark concerning II:} The definition of $\dr$ in part I of the theorem
involved a subspace
$W^*$ of $V^*$; in part II we have defined $W^*$
abstractly rather than as a subspace of $V^*$, so it is $\tau(W^*)$,
which is isomorphic to $W^*$ but {\em is} a subspace of $V^*$, to
which we restrict states in defining $\dr$.  (Of course, $W^*$ {\em
itself} is a subspace of $V^*$ according to the category-theoretic
definition of subspace.)

{\bf Proof:}   To prove part I, we must show that
$D$ is a cone in $W$, and $\lambda$ separates it from $-D$.
It is easy to verify linearity of $\dr \equiv \pi$ from the
definition, and in finite dimensions, it is also easy to verify that
linear maps from one vector space {\em onto} another (such as $\dr$)
take  cones to  cones.  For all $x \in \dot{C} =: C - \{0\}$,
$\lambda[x] > 0$.  But $\lambda[x] = x[\lambda]$ by duality, and by
the definition of $\dr$ and the fact that $\lambda \in W^*$,
$x[\lambda] = x\dr [\lambda] \equiv \lambda[ x \dr]$, so $\lambda[x\dr]
> 0$ for all $x \in \dot{C}$, i.e. (since $\dr$ maps $\dot{C}$ onto
$\dot{D}$), $\lambda[y] > 0$ for all $y \in \dot{D}$.  That is,
$\lambda$ separates $D$ from $-D$.  

To prove part II, let $\tau$ be $\pi^*$.  That is, for all $x \in W^*$,
$\eta \in V, \tau(x)[\eta] = x[\pi(\eta)]$. By duality,
this gives $\eta[\tau(x)] = \pi(\eta)[x]$. Since, by
Proposition~\ref{epimono}, $\tau$ is an injection, this last equation
determines $\pi(\eta)$ to be essentially $\eta\dr_{\tau(W^*)}$, as
desired.  The ``essentially'' refers to the fact that $\pi(\eta)$ is
actually the pullback along $\tau$ of this restriction; the two are
the same function only if one identifies $W^*$ with its image under
$\tau$.  $\tau$, in other words, tells us how $W^*$ can be identified
with a subspace of the full space $V^*$ of observables, in such a way
that $\pi(\eta)$ becomes identified with the restriction of $\eta$ to
$W^*$. $\Box$

\begin{proposition}\label{prop: basic preimage facts}
In a cone-pair, $\pi$ has the property that
for $x \in \extr{\hat{D}}$, the set $\pi^{-1}(x)$ is convex, compact,
and closed, and its extremal elements are extremal in $\hat{C}$.
\end{proposition}

{\bf Proof:} Convexity is immediate: if $y_1, y_2 \in C$, $\pi(y_1)=
x$ and $\pi(y_2) = x$, then $\lambda y_1 + (1 - \lambda) y_2 \in C$ by
convexity of $C$, and by linearity of $\pi$, $\pi(\lambda y_1 + (1 -
\lambda) y_2) = \lambda(\pi(y_1)) + (1 - \lambda) \pi(y_2) = x$.
Closedness of $\pi^{-1}(x)$ in the Euclidean metric topology 
follows from the fact that $\pi$, being a
function from a finite-dimensional inner product space to a
finite-dimensional normed space, is continuous (cf. e.g.
\cite{Young88a}, Exercise 7.3), and the preimage of a closed set under
a continuous function is closed (cf. e.g. \cite{Kripke68a}, Corollary
IV.C.4).  Since finite intersections of closed sets are closed, $C
\cap \pi^{-1}(x)$ is closed as well.  Compactness follows from the
fact that $\hat{C}$ is compact (cf. e.g.  \cite{Barvinok2002a}) hence
a compact metric space, and a closed subset of a compact metric space
is compact (\cite{Kripke68a}, Corollary VII.A.11).

Now let $x \in \extr{\hat{D}}$, and let $y \in \pi^{-1}(x)
\cap C$ not be extremal in $\hat{C}$.  We need to show that such a $y$ is
not extremal in $\pi^{-1}(x)$ either.  $y \notin \extr{\hat{C}}$ means
there are $y_1, y_2 \in \hat{C}$ with $y_1 \ne y_2$, $y = \lambda y_1 + (1 -
\lambda)y_2$.  By linearity of $\pi$, $x \equiv \pi(y) = \lambda
\pi(y_1) + (1 - \lambda) \pi(y_2)$; since $x \in \extr(\hat{D})$,
$\pi(y_1) = \pi(y_2) = x$.  Hence $y_1, y_2 \in \pi^{-1}(x)$, so $y
\notin \extr{(\pi^{-1}(x) \cap C}$.  $\Box$

In important classes of examples, a stronger property holds:

\begin{definition} \label{def: unique preimage property}
A cone-pair including $C,D,\lambda,\pi$ is said to have the {\em
unique preimage property} (UPIP) if $x \in \extr{\hat{D}}$ implies
that $\pi^{-1}(x)$ consists of a single element (which must therefore
be extremal).
\end{definition}

Equivalently (because of Prop. \ref{prop: basic preimage facts}),
extremal reduced states have only extremal preimages.

\begin{problem} \label{prob: when UPIP}
Find nontrivial necessary and/or sufficient conditions (some are given
below, but others almost certainly exist) for cone-pairs $C, D, \pi$
to have the UPIP.
\end{problem}

Finally, note that the converse of the UPIP follows from Proposition 
\ref{prop: basic preimage facts}:  If $\pi^{-1}(x)$ is unique, 
then it must be extremal, and $x$ must be extremal as well.

\large
\section{GENERALIZED ENTANGLEMENT IN SPECIAL CLASSES OF CONES}
\normalsize  

We now formally define several ``settings'' in which to study GE;
these are special classes of cone-pairs, physically and/or
mathematically motivated.

\begin{definition}
\begin{mlist}
\item {\em Distinguished quantum observables setting}, defined above.
An equivalent formulation is the {\em Hermitian-closed (aka
$\dagger$-closed) operator subspace setting}, in which the
distinguished observable subspace is the Hermitian operators belonging
to a $\dagger$-closed subspace, containing the identity operator, of
the complex vector space of all linear operators on a quantum system.
\item {\em Lie-algebraic setting.}  
Here, $C$ is the cone of positive Hermitian operators on a
(finite-dimensional) Hilbert space carrying a Hermitian-closed Lie
algebra $\fg$ (playing the role of $W^*$) of Hermitian operators (with
Lie bracket $[X,Y] := i(XY-YX)$, and containing the identity operator)
and $D$ the cone (in $(W^*)^* =: W$) of linear functionals on $\fg$
induced from positive Hermitian elements of $C$ by restriction to
$W^*$.
\item {\em Associative algebraic setting.} Here, the distinguished
observables are the Hermitian elements of some associative subalgebra
of the associative algebra of all operators on a quantum system.
\end{mlist}
\end{definition}

By construction, the Lie-algebraic and associative algebraic settings
are special cases of the distinguished quantum observables case.  As
noted in ~\cite{BKOV2002a},
since the Lie-algebraic setting was defined to involve
finite-dimensional $\dagger$-closed matrix representations, the Lie
algebras involved are necessarily reductive i.e., the
direct product\footnote{As Lie algebras; the induced direct product of
the algebras considered as vector spaces (i.e. without their Lie
bracket structure) is also a vector space direct sum.} of a semisimple
and an Abelian part.

A distinction that can be nontrivially made within all the settings in
the above list is between those in which the distinguished observables
act {\em irreducibly}, and those in which there is a nontrivial
subspace invariant under the action of all observables.

\begin{proposition}
In the $\dagger$-closed operator subspace setting, the distinguished
subspace has a basis of Hermitian operators that is orthonormal in the
trace inner product $\langle A, B \rangle = \tr AB$.
\end{proposition}

Because of this proposition, we may construct an orthogonal projection
operator (some would call it a superoperator) $\Pi_S$, acting on the
space of Hermitian operators by projecting into the subspace of
distinguished observables.  We can also use such a basis to define a
measure of entanglement for pure states, the {\em relative purity}
(although the name may be slightly misleading, for reasons we will explain).

\begin{definition}  
Let $\omega$ be a state on a $\dagger$-closed set $S$ of quantum
observables.  The {\em purity} $P(\omega)$ of a state $\omega$ is
defined by letting $X_\alpha$ be an orthonormal (in trace inner
product) basis of $S$.  Then \beqa P(\omega) := \sum_{\alpha}
(\omega(X_\alpha))^2\;.  \eeqa
\end{definition}
%We also use variants of the purity where the common normalization
%constant of the orthonormal basis for $S$ is chosen differently, for
%instance, so that the maximum value of $P(\omega)$ is unity.
Note that any state $\omega$ on the {\em full} operator space
corresponds uniquely to a density operator $\rho_{\omega}$, defined by the
condition $\tr (\rho_\omega X) = \omega(X)$ for all observables $X$.

Closely related to the above purity is the {\em relative purity} of a
pure state $\ket{\psi}$ of the overall quantum system; this is defined
equal to the purity of the state it induces on $S$, or equivalently,
with $X_\alpha$ as above, \beqa P_S(\ket{\psi}) := \sum_{\alpha \in S}
|\dmelement{\psi}{X_\alpha}|^2\;.  \eeqa In fact, this definition
could be straightforwardly extended to mixed states $\omega$ on the
full Hilbert space, as \beq P_S(\omega):= \sum_{\alpha \in S} |{\rm tr~}
\omega X_\alpha|^2\;.  \eeq
However, a requirement for entanglement measures is
convexity~\citep{Vidal2000b}, and the above extension lacks this as
well as other desirable properties.  We will generally extend
pure-state entanglement measures $\mu$ to mixed states via the {\em
convex hull (often called convex roof) construction} 
(cf. e.g. \cite{Uhlmann98a, Bennett96a})
standard in ordinary entanglement theory: the value of the measure 
$\mu$ on
a mixed state $\omega$ is the infimum, over convex decompositions
$\omega = \sum_i p_i \pi_i$ of $\omega$ into pure states $\pi_i$, of
the average value of the pure-state measure, that is, of $\sum_i p_i
\mu(\pi_i)$.  This is convex by construction.

Defining $\Pi_S$ as the projection superoperator onto the operator
subspace $S$, it is easily verified that \beqa P_S(\omega) :=
\sum_{\alpha} |\tr \Pi_S(\rho_{\omega}) X_\alpha|^2
\equiv \tr
[\Pi_S(\rho_\omega)^2]\;.  \eeqa
For any density operator $\rho$, we call $\Pi_S(\rho)$ the associated
{\em reduced} density operator; note that it need {\em not} be a
positive operator on the full state space (although it is in the
standard multipartite case).  This is not problematic because for any
PSD element $R$ of the {\em distinguished} observable space, $\tr \Pi_S(\rho)
R \ge 0$, of course.

The following proposition is immediate from Theorem 14 of~\cite{BKOV2002a}.

\begin{proposition} \label{prop: mre implies ge in irreducible lie}
In the irreducible Lie-algebraic setting, pure states with maximal
relative purity are generalized unentangled.
\end{proposition}

The converse is {\em not} true in general.   Also, 
the analogue of  Prop. \ref{prop: mre implies ge in irreducible lie}
for the general Lie-algebraic setting (allowing {\em reducible}
representations) can be shown by example to be false.

Another situation in which maximal relative purity implies generalized
unentanglement is embodied in the following.

\begin{proposition} \label{prop: when else mrp implies ge}
In the $\dagger$-closed operator subspace setting states with unit
relative purity have unique preimages, and are 
therefore generalized unentangled.
\end{proposition}

\noindent
{\bf Proof:} A necessary and sufficient condition for a normalized
state $\omega$ on the space of all observables to be pure is $\tr
(\rho_\omega^2) = 1$.  (Henceforth we suppress the $\omega$-dependence
of $\rho$.)  Letting $X_\alpha$ be an orthonormal basis for the space
of all observables such that a subset (denoted by the letter $\beta$
for the index) indexes the distinguished subspace $S$, with another
subset (indexed by $\gamma$) indexing $S^\perp$, and writing $\langle X_\alpha
\rangle$ for $\tr \rho X_\alpha$, we have $\rho = \sum_\alpha
\langle X_\alpha \rangle X_\alpha$.
From this and orthonormality of the $X_\alpha$ it is easy to see that
$\tr (\rho^2) = \sum_\alpha \langle X_\alpha \rangle^2$.  
$P_S(\rho) \equiv \sum_{\beta \in S} \langle X_\beta \rangle^2$; since
extremal overall states have $\tr (\rho^2) = 1$, $P_S(\rho)$ for
a pure state $\rho$ can never be greater than $1$, since it is a sum
of a subset of the positive quantities $\langle X_\alpha \rangle^2$ which 
sum to $1$.
Let $X$ have unit relative purity, i.e. $\sum_{\beta \in S} X_\beta^2 = 1$.
This implies $\sum_{\gamma \in S^\perp} \langle Y_\gamma \rangle^2 =
0,$ which requires $\langle X_\gamma \rangle = 0$ for all $\gamma \in
S^\perp$.  Thus, $P_S(\rho)$ has a unique preimage, namely itself, so
$\rho = P_S(\rho)$.  If $P_S(\rho)$ did not induce an extremal state
in the convex set of reduced states, it would be a convex combination
of distinct operators $\rho_1$ and $\rho_2$ inducing distinct reduced
states; these would have distinct preimages, but the convex
combination of these preimages be $P_S(\rho) \equiv
\rho$, violating the assumption that $\omega$ is pure. $\Box$

%Note that the converse of this statement is {\em not} true in general,
%although (as shown in the proof of the preceding proposition), it is
%immediate when the UPIP holds.

What about states whose relative
purity is maximal among all states, even when this maximal value is
not unity?  When the maximum is not unity, no pure state has an
unchanged reduced density matrix: all pure state density
matrices project to reduced ``density matrices'' that are either
mixed, or not even PSD.  Thus we cannot immediately
conclude that $\sum_\beta \langle X_\beta \rangle^2 =1$, so we do not
have $\langle X_\gamma \rangle = 0 $ for all $X_\gamma \in S^\perp$.
If there is nevertheless a unique preimage, i.e. the $X_\gamma$ are
uniquely determined by the $X_\beta$ (for $\beta$ indexing $S$), it
must be a consequence of positive semidefiniteness of the initial
state, since linear algebra alone gives no restrictions on the
$X_\gamma$.
However, because relative purity is a strictly convex function 
of the reduced density matrix, a state's having 
maximal, even if not unit, relative purity, 
implies generalized unentanglement in the $\dagger$-closed
operator subspace framework.  It does not, however, imply the other
part of Proposition \ref{prop: when else mrp implies ge}, that the 
reduced state has a unique preimage.  Formally:

\begin{proposition} \label{prop: max purity implies GE}
Let $x \in \hat{C}$ be such that the relative purity of $x$ is no less
than that of every other element of $\hat{C}$.  Then $x$ is generalized
unentangled.
\end{proposition}

\noindent
{\bf Proof:}
The relative purity of $\omega$ is just the Euclidean norm of $\Pi_S(\rho_\omega)$ (with
respect to the trace inner product).  Suppose $\omega$ has maximal relative
purity, i.e. $|\Pi_S(\rho_\sigma)|| \le ||\Pi_S(\rho_\omega)||$ 
for all $\sigma \in \hat{C}$.  Suppose
there are $\alpha, \beta \in \hat{D}, \alpha \ne
\beta,$
such that $\Pi_S(\rho_\omega) = \mu \rho_\alpha + (1 - \mu) \rho_\omega$.  
Then by the triangle inequality
$||\Pi_S(\rho_\omega)|| \le 
||\mu \rho_\alpha|| + ||(1 - \mu) \rho_\beta|| = \mu ||\rho_\alpha || + (1 - \mu) 
||\rho_\beta||$.
Since neither $||\rho_\alpha ||$ nor $||\rho_\beta||$ is greater than 
$||\Pi_S(\rho_\omega)||$, we must 
have $||\rho_\alpha|| = ||\rho_\omega|| = ||\Pi_S(\rho_\omega)||$, 
so there is equality in the triangle 
inequality.  That requires $\mu \rho_\alpha$ to be proportional to $(1 - \mu) \rho_\beta$,
however, so that $\rho_\alpha = \rho_\beta = \Pi_s(\rho_\omega)$.  
This shows that $\Pi_S(\rho_\omega)$ is extremal in the set of reduced density operators
corresponding to states in  
$\hat{D}$.  In other words, $\omega$ is generalized unentangled. $\Box$

%\begin{proposition} \label{unique preimage implies GE have maximal purity}
%In the $\dagger$-closed operator subspace setting when the UPI
%property holds, generalized unentangled states have maximal relative
% purity.
%\end{proposition}

%\begin{problem} Is the converse of Proposition 
%\ref{unique preimage implies GE have maximal purity} true?  That is,
%is it the case that when in a given instance of the $\dagger$-closed
%operator subspace setting, every generalized unentangled state has
%maximal relative  purity, that instance has the UPI
%property?
%\end{problem}

%We can also ask whether a yet stronger statement holds:

%\begin{problem}
%In the $\dagger$-closed operator subspace setting, does every
%$\pi$-image of a generalized unentangled state that has maximal
%relative  purity have a unique preimage?
%\end{problem}

It follows from the representation theory of associative
algebras that the UPIP holds for the irreducible associative
algebraic setting.  The other case in which we know it holds is the
irreducible semisimple Lie algebraic setting.  In this setting, the
observables consist of the Hermitian part (itself a real Lie algebra)
of a complex Lie algebra represented faithfully and irreducibly by
matrices acting on a finite-dimensional complex Hilbert space, and
including the identity matrix $I$.  Such Hermitian parts of
irreducible matrix Lie algebras are precisely the real semisimple
algebras possibly extended by the identity.  The identity is
relatively unimportant since all normalized states have the same value
on it: the normalization condition is the affine plane $\omega(I)=1$,
so the convex structure of the state space is entirely determined by
the expectation values of the traceless operators.  We introduce a bit
more notation in order to state a result, proved in~\cite{BKOV2002a},
that includes this and other important facts about the irreducible
Lie-algebraic case.

A \emph{real} Lie algebra of
Hermitian operators may be thought of as a distinguished
family of Hamiltonians, which generate (via $h \mapsto e^{ih}$) a Lie
group of unitary operators, describing a distinguished class of
reversible quantum dynamics.  
More generally, we might want
Lie-algebraically distinguished completely positive (CP) maps, $\rho
\mapsto \sum_i A_i \rho A_i^\dagger$ describing open-system quantum dynamics.  We call
the operators $A_i$ the ``Hellwig-Kraus'' or ``HK'' operators, since
they appear to have been introduced in \cite{Hellwig69a,Hellwig70a}
(see also \cite{Kraus83a, Choi75a}).  The HK operators for a given CP
map are not unique, but this does not lead to nonuniqueness of any of
the objects we define in terms of them below.  A natural Lie-algebraic
class of CP-maps has HK operators $A_i$ in the
topological closure $\overline{e^{\fh_c\oplus \one}}$ of the Lie group
generated by the \emph{complex} Lie algebra
$\fh_c\oplus\one$~\footnote{ $\fh_c$ is constructed by taking the
complex linear span of a basis for $\fh$. $\fh_c \oplus \openone$
guarantees inclusion of the identity operator $\openone$.}.  Having HK operators in a
group ensures closure under composition.  Using $\fh_c \oplus \one$
allows non-unitary HK operators.  Topological closure introduces
singular operators such as projectors.  
\iffalse
The following theorem (Theorem
14 in \cite{BKOV2002a}) characterizing GUE states
shows the power of the Lie-algebraic setting.  
\fi

Define an $\fh$-state to be a linear functional on a {\em complex}
matrix Lie algebra $\fh$ belonging to the convex set of such states induced
by normalized quantum states on the full representation space.
Complex-linearity ensures that the convex structure of such a state
space is the same as that of the states induced by taking as the
distinguished observables only the Hermitian elements (a real Lie
algebra we denote $\real({\lie{h}})$), which is the definition we used
above for the Lie-algebraic setting.

\newcounter{stcharcnt}
\begin{theorem} 
\label{thm:stchar}
Let $\fh$ be a complex irreducible matrix Lie algebra, 
with $\trless{\lie{h}}$ its traceless (semisimple) part
and $\real({\lie{\fh}})$ its Hermitian part. 
The following are equivalent for a density matrix $\rho$
inducing the $\lie{h}$-state $\lambda$:
\begin{mlist}
  \refstepcounter{stcharcnt}\label{st1}
  \item[\emph{(\textbf{\thestcharcnt})}] $\lambda$ is a pure
  $\lie{h}$-state, that is, it is extremal in the convex set of
  normalized linear functionals on $\lie{h}$.
  \refstepcounter{stcharcnt}\label{st2}
  \item[\emph{(\textbf{\thestcharcnt})}] $\rho=\ketbra{\psi}{\psi}$ with 
  $\ket{\psi}$ the unique ground state of 
  some $H$ in $\real(\lie{h})$.
  \refstepcounter{stcharcnt}\label{st3}
  \item[\emph{(\textbf{\thestcharcnt})}] $\rho=\ketbra{\psi}{\psi}$ with
  $\ket{\psi}$ a minimum-weight vector (for some simple root system
  of some Cartan subalgebra) of $\trless{\lie{h}}$.
  \refstepcounter{stcharcnt}\label{st5}
  \item[\emph{(\textbf{\thestcharcnt})}] $\lambda$ has maximum purity
  relative to the subspace $\real(\lie{h})$ of observables.
  \refstepcounter{stcharcnt}\label{st6}
  \item[\emph{(\textbf{\thestcharcnt})}] $\rho$ is a one-dimensional
   projector in the topological closure of $\overline{e^{\lie{h}}}$.
\end{mlist}
\end{theorem}
%% \ignore{
%%   \refstepcounter{stcharcnt}\label{st4}
%%   \item[\emph{(\textbf{\thestcharcnt})}] $\rho=\ketbra{\psi}{\psi}$ such that
%%   $R=\{r\in\trless{\lie{h}} \suchthat r\ket{\psi}=\lambda_r\ket{\psi}\}$ 
%%   satisfies $R+R^\cstar=\trless{\lie{h}}$.
%% }
\begin{problem} \label{prob: when GE implies mrp}
Does the implication from GUE to maximal relative
purity, hold in other natural situations?
\end{problem}

As already noted it is fairly easy to show by example that in the
Lie-algebraic setting but without the assumption of irreducibility,
the UPIP need {\em not} hold.  A more general question
suggests itself:

\begin{problem} \label{prob: irreducible implies UPIP}
In the $\dagger$-closed operator subspace setting, does the UPIP
hold whenever the distinguished operators act irreducibly?
\end{problem}

\large
\section{ANALOGUES OF LOCAL MAPS}
\normalsize

Our work on GE raises many questions arising from the closely
related problems of finding natural generalizations or
analogues of the notions of LOCC ({\em Local Operations and Classical
Communication}) and of monotone entanglement measures (or {\em
entanglement monotones} \citep{Vidal2000b}).  The relation comes from
requiring that a reasonable entanglement measure be nonincreasing
under LOCC operations; if one found a natural generalization of this
notion of LOCC to our more general settings, it would also be natural
to look for measures of GE monotone under this generalization.  
Here, we briefly present some ideas from \cite{BKOV2002a} (with 
a few minor extensions) on how to generalize LOCC; 
that paper contains more on this topic and on 
GE measures.  Some of the most fundamental questions remain open, 
so we will concentrate on sketching the situation in hopes of
stimulating further work.

The semigroup of LOCC maps, introduced in~\cite{Bennett96c}, and the
preordering it induces on states according to whether or not a given
state can be transformed to another by an LOCC operation are at the
core of entanglement theory.  LOCC maps are precisely those
implementable by using CP quantum maps on the local subsystems, and
classical communication, e.g.  of ``measurement results,'' between
systems.  
We now formalize this notion, beginning with the notion 
of {\em explicitly decomposed} map which, however, can apply to the general
case, not just the quantum one.
An {\em explicitly decomposed} trace-preserving map
$\{M_k\}_{k \in K}$ is a set of maps $M_k$ that sum to a
trace-preserving one $M$.  The {\em conditional composition} of an
explicitly decomposed map $\{M_k\}_{k \in K}$ with a set of explicitly
decomposed maps $N_k := \{N_{nk}\}_{n \in N_k}$ is the explicitly
decomposed map $\{ N_{nk} \circ M_k \}_{k \in K, n \in N_k}$.  We can
view each $M_k$ as being associated with measurement outcome $k$,
obtained (given a state $\omega$) with probability 
$\tr M_k (\omega)$, and
leading  to the state $ M_k (\omega)$ when outcome $k$ is obtained.  The
conditional composition of $\{M_k\}_{k \in K}$ and $\{N_{nk}\}_{n \in
N_k}$ can be implemented by first applying $M$ and then, given
measurement outcome $k$, applying $N_k$.  There are analogous definitions
of explicitly decomposed maps and conditional composition
without the trace-preservation
condition.  

In the usual quantum case, closing the set of one-party (aka {\em unilocal})
maps (for all parties) under conditional composition gives the LOCC
maps.
The semigroup generated by composition of unilocal
explicitly decomposed maps having a single HK operator in their
decomposition, is often known as SLOCC (for {\em stochastic} LOCC).
SLOCC involves local quantum
measurements and classical communication conditional on a {\em single}
sequence of local measurement results, when each local measurement is
performed in a manner that preserves all pure states (i.e., with a
single HK operator for each outcome).  Its mathematical structure is
relatively simple, as the part generated by
nonsingular HK operators is the trace-nonincreasing part of a
representation of a product of various GL$(d_i)$, with the factors
acting on local systems of dimension $d_i$~\footnote{We are not
certain if the full LOCC semigroup is the trace-nonincreasing part of
the topological closure of this representation, but it seems a
reasonable possibility.}.

When the distinguished observables form a semisimple Lie algebra
$\fh$, a natural multipartite structure can be exploited to generalize
LOCC, as well as the larger, more tractable class of 
{\em separable} maps; see \cite{BKOSV2004a,BKOV2002a}. 
In generalizing
LOCC to the convex setting, two aspects of LOCC must be considered: first, that it
constrains maps to be {\em completely positive}; second,
that it also constrains them to have certain {\em locality} properties.

%\subsection{Towards Generalizing Complete Positivity}

A \emph{positive} map of $D$ is a linear map ${A}:V\rightarrow V$ such
that ${A}(D)\subseteq D$. The map ${A}$ is \emph{trace preserving} if
$\trace(x)=\trace({A}(x))$ for all $x$. This definition corresponds to
positive, but not necessarily CP, maps in quantum
settings.  Without additional algebraic structure, 
it is not possible to define a unique ``tensor product'' of
cones, as would be required to distinguish between positive and
CP maps~(\cite{Namioka69a,Wittstock74a} (cited
in~\cite{Wilce92b})).
In a continuum of possible products of cones, there are two natural
possibilities that are in a sense the two extremes.  The first is the
convex closure of the set of tensor-products of the cones' vectors,
which for the case of the product of two quantum systems' unnormalized
state spaces gives the separable states of the bipartite system. The
second is to use the dual cone of the cone obtained by applying the
first construction to the duals of the cones; in the quantum case, it
gives the set of (unnormalized) states that are positive on product
effects (this is isomorphic to the cone of positive but not CP
operators between the state spaces, by the ``Choi-Jamiolkowski''
isomorphism between $V \otimes V$ and $\cl(V)$).  It is not clear how
to pick out a natural case between these extremes in general without
adding algebraic structure, except perhaps if the cones are self-dual
with respect to non-degenerate inner products on the real vector
spaces. In that case, one could pick a self-dual cone between the two
constructions (which would give the usual state space of a bipartite
system in the quantum case).

The family of positive maps of $C$ is closed under positive
combinations and hence forms a cone. In the Lie-algebraic, or even the
bipartite setting, the extreme points of this cone are not easy to
characterize (see, for example,~\cite{Wilce92b},
p. 1927,~\cite{Gurvits2002a}).  We seek generalizations of 
the notion of complete positivity to the cones setting.  
We might explicitly introduce a cone representing the
``tensor product'' extension of $D$ and require extendibility or
``liftability'' of the map to $D$.  Another, perhaps more uniquely
determined, approach might begin from the observation that the extreme
points of the cone of completely positive maps are extremality preserving:
for all extremal (belonging to an extreme ray) $x\in D$,
${A}(x)$ is extremal.  However there are extremality preserving
positive, not CP, maps.  An example is partial transposition for
density operators of qubits.  In \cite{BKOV2002a} we explore how one
might rule these out.  There is also the question of why extremality
preservation would be a natural physical or operational, as opposed
to mathematical, requirement.  

%\subsection{Towards Generalizing Locality}
%Another approach to GLOCC is
%to hope that the restriction to CP maps might either
%emerge, or be imposed, late in the game, concentrating instead on
%generalizing locality.  
\iffalse
 one runs the risk of ending
up with, say, locally {\em positive} but not necessarily completely
positive maps, and classical communication, so one must hope that in
such a case one will see how to exclude non-CP maps by additional
natural requirements.

The next step is to define a family of maps that generalizes the
separable maps.  Call a positive map ${A}$ of $D$ $C$-separable if it
is a mixture of extremality-preserving positive maps ${A}_k$ that are
also extremality-preserving and positive for $D_{\textrm{\tiny sep}}$.
In the bipartite setting, this definition includes maps such as the
swap, which exchanges the two subsystems and is not separable, in
addition to some non-completely positive operations. Note that if the
Lie-algebraic definition of separability is used, operations like the
swap are excluded because they are not in the Lie group generated by
$\lie{h}_l$: The swap induces an exterior automorphism of
$\lie{h}_l$. From the point of view of entanglement, including the
swap can make sense because it obviously does not increase
entanglement.
\fi
To try to generalize the notion of locality, we introduce the idea of
{\em liftability}.  We say that a positive map ${A}$ on $D$ can be
lifted to $C$ if ${A}$ preserves the nullspace of $\pi$, or,
equivalently, if there exists a positive map ${A}'$ on $C$ such that
$\pi({A}(x))={A}'(\pi(x))$. In this case, we say that ${A}'$ is the
lifting of ${A}$ to $C$.

In standard multipartite quantum entanglement, {\em
unilocal} maps (ones that act nontrivially only on one factor) are
liftable to the cone of local observables; they have a
well-defined action there.  But so are tensor product maps $\ca
\otimes \cb \otimes \cdots \cz$, and in the case when some of the
subsystems are of the same dimension, so are maps performing
permutations among the isodimensional factors.  To get LOCC we would
need to rule out the latter two cases, leaving the unilocal maps; then
one can generate a semigroup from the unilocal maps by conditional
composition of explicitly decomposed trace-preserving maps.  On the
other hand, in the standard quantum case the semigroup of maps generated by
conditional composition of maps liftable to the distinguished subcone
might enjoy many of the same properties of the usual LOCC maps, so it may 
be worth study in the general setting.

\begin{problem} \label{abcd}
Is the semigroup generated by {\em completely positive} unilocal quantum
maps and pairwise exchanges of isodimensional systems the full
semigroup generated by conditional composition of
liftable-to-local-observables explicitly decomposed maps?  
%Does it give rise to the same partial ordering of states as
%standard ``SLOCC''?
\end{problem}

Note that using liftability to define locality may be of some help in
ruling out local non-completely positive maps, since all maps must be
positive on the overall cone.  It is especially helpful if the answer to 
Problem \ref{abcd} is ``yes.''
When no subsystem has dimension greater than the square root of 
the overall dimension, it is then fully effective
in imposing complete positivity, because for any local map $M$, 
complete positivity of $M$ is equivalent to positivity of the unilocal
map $\Id \otimes M$ where the identity map $\Id$ acts on a Hilbert space at least as 
large as the one $M$ acts on.

In the standard multipartite quantum
case, the high degeneracy of unilocal operators can also be used to
help distinguish them in a way not so directly dependent on explicit
introduction of cones to represent individual systems---and similarly
one can use spectral information about HK operators to characterize
ones that act on the {\em same} single system, thereby characterizing
LOCC in terms of conditional composition of explicitly decomposed maps
whose HK operators together satisfy certain spectral conditions
\citep{BKOV2002a}. However, it is not
clear how to abstract this to general cones.  Perhaps
something can be done with the facial structure of the
cone $D$, or of the cone of positive maps on $D$ (or of other
subcones of maps chosen as abstractions capturing aspects of complete
positivity).  A more in-depth investigation of dynamics generalizing
LOCC thus remains as a challenging and many-faceted area for research,
as does the investigation of measures of GE nonincreasing under such
maps.

{\bf Acknowledgements} We thank Manny Knill for valuable discussions
and for collaboration on earlier work summarized and built upon in the
present paper.  Work at Los Alamos was supported by the US DOE through
Los Alamos National Laboratory's Laboratory Directed Research and
Development (LDRD) program.  

%\large
%\bibliography{entcones}
%\normalsize

 \end{document}